\documentclass[10pt]{article}

\usepackage{amsmath}

\usepackage{amssymb}

\usepackage{graphicx}

\usepackage{cite}

\usepackage[usenames]{color}
\usepackage{cancel}
\usepackage[normalem]{ulem}

\usepackage[labelfont=bf,labelsep=period,justification=raggedright]{caption}

\makeatletter
\renewcommand{\@biblabel}[1]{\quad#1.}
\makeatother

\date{}

\begin{document}

\begin{flushleft}
{\Large
\textbf{A Comment on ``Cycles and Instability in a Rock-Paper-Scissors Population Game: A Continuous Time Experiment"}
}
\\
 {Zhijian Wang$^{1,\ast}$}
Siqian Zhu$^{1,2}$,
Bin Xu$^{3,1,4}$,
\\
 {1} Experimental Social Science Laboratory, Zhejiang University. \\
 {2} Chu Kochen Honors College, Zhejiang University. \\
 {3} Public Administration College, Zhejiang Gongshang University. \\
 {4} SKLTP, ITP, Chinese Academy of Sciences. \\
 {*}correspondence: wangzj@zju.edu.cn
\end{flushleft}

\section*{Abstract}
The authors (Cason, Friedman and Hopkins,
\emph{Review of Economic Studies}, 2014) claimed a conclusion
that the treatments (using simultaneous matching in discrete
time) replicate previous results that exhibit ¡°weak or no cycles¡±.  After
correct two mathematical mistakes in their cycles tripwire algorithm,
we research the cycles by scanning the tripwire in the full strategy space of the games and we find significant cycles missed by the authors. So we suggest that, all of the treatments exhibit significant cycles.
\begin{flushleft}
\textbf{JEL numbers}: C72, C73, C92, D83; \\
\textbf{Keywords}: experiments, learning, cycles, mixed equilibrium, discrete time
\end{flushleft}
~\\
The existence of cycles in mixed equilibrium games is a cutting edge question in the field
crossing game theory \cite{kuhn1997work} and evolutionary game theory \cite{Maynard1982evolution} for decades.
The finding of cycles's existences by quantitative measurement in controlled experiments reported by Cason, Friedman and Hopkins
\cite{Friedman2012} is a milestone-like contribution in the field.
But one of the  conclusions in their article attracts our attention.

In their article \cite{Friedman2012},
they claimed that
control treatments (using simultaneous matching in
discrete time) replicate previous results that exhibit \emph{``weak or no cycles"}.
This depended on the results ------ The stable discrete time  treatments
 (SD) do not exhibit clear cyclical behavior indicated by
 \emph{CRI's not significantly different from 0} (see their Table 3).

 Two mathematical mistakes in the measurement algorithm were found when we checked
 their results above.\footnote{We thank the authors' \cite{Friedman2012}
 confirmation on this point during ESA-NA (2014) Conference.}
  We corrected the mistakes (see Appendix).
 Using the refined measurement,
 if setting the start point $(\alpha,\beta)$ of the tripwire for counting cycles (see Fig.~\ref{fig:tripwire})
at ($\frac{1}{4}, \frac{1}{4}$) ------ Nash equilibrim (NE) of the games as \cite{Friedman2012},
CRI of SD will still not be significantly different from 0  (see up panel in Table 1 comparing with their Table 3).
  So, for SD,
  we agree that,
  there are only weak or no cycles \emph{around NE}.

However, 
if we  set $(\alpha,\beta)$
at (0.23, 0.26) for SD-Mixed and
at (0.22, 0.40) for SD-Pure,
 using the CRI measurement as criterion still,\footnote{ 
Our pilot results suggest that using angular motion of the transits (e.g. the $\theta$ in Eq.(10) in \cite{2013arXiv1301.3238X}) are efficient observation for cycles too. We wish to return to these future.
}
we can find, both CRI of SD are \emph{significantly different from 0}  which indicates cycles' existence (see low panel in Table 1).


  Meanwhile, if using the accumulated counting number of cycles $C$ as index (see the Eq.(2) in \cite{2013arXiv1301.3238X} for detail explanation) instead of CRI,
we also find that all $\bar{C}$ (of the experimental blocks) for the treatments are significantly different from 0 (see Table~2). This, again, indicates the cycles' existence. 

 So, for SD, we suspect that their conclusion of \emph{``weak or no cycle"}.  We suggest that the treatments (using simultaneous matching in discrete time) exhibit significant cycles instead of 
\emph{``weak or no cycles"}.\footnote{We would like to point out follows. (1) The $\alpha,\beta$ for the two SD is close to the actual mean observation of the aggregated social strategy which seems to tell us that, cycles are actually rounding  actual mean observation instead of NE. (2) Their
 treatments of  RPS games are economical because only 6 groups of 8 subjects and 15
 minutes are needed. If cycle could be obtained, such treatments could be  exemplified
 classroom experiments for teaching evolutionary game theory.
}
  
~\\

\begin{table}[h]
\center
\small
\begin{tabular}{|lccccc|c}
\hline
 &&\emph{Number of}	    &\emph{Number of }	&\emph{Cycle }	& \\
\emph{Game Condition}	&($\alpha, \beta$)&\emph{Counter-Clock}	&\emph{Clockwise}	&\emph{Rotation}	
&~~\emph{p}-value\\
                	&&\emph{-wise Transits}          &\emph{ Transits}	   &\emph{Index} (CRI)	& \\
\hline
$S$ Continuous-Instant	&($\frac{1}{4}$, $\frac{1}{4}$)   &25.2	&5.5	&~0.65\textbf{*}	    &0.028\\
$S$ Continuous-Slow	    &($\frac{1}{4}$, $\frac{1}{4}$)   &9.4	&0.8	&~0.87\textbf{*}	    &0.027\\
$S$ Discrete-Mixed (SD-Mixed)	     &($\frac{1}{4}$, $\frac{1}{4}$)  &2.3	&1.2	&~0.38~~	    &0.116\\ 
$S$ Discrete-Pure ~~(SD-Pure)  &($\frac{1}{4}$, $\frac{1}{4}$)  &1.0	&1.0	&~0.13~~	    &0.753\\ 
$U_a$ Continuous-Instant&($\frac{1}{4}$, $\frac{1}{4}$)   &31.9	&0.9	&~0.94\textbf{*}	&0.027\\
$U_a$ Continuous-Slow	&($\frac{1}{4}$, $\frac{1}{4}$)   &8.2	&0.0	&~1.00\textbf{*}	&0.014\\
$U_a$ Discrete-Mixed	&($\frac{1}{4}$, $\frac{1}{4}$)   &2.3	&0.2	&~0.79\textbf{*}	&0.027\\
$U_a$ Discrete-Pure	    &($\frac{1}{4}$, $\frac{1}{4}$)   &1.9	&0.3	&~0.79\textbf{*}	&0.027\\
$U_b$ Continuous-Slow	&($\frac{1}{4}$, $\frac{1}{4}$)   &0.3	&8.5	&-0.93\textbf{*}	&0.028\\
\hline
$S$ Discrete-Mixed (SD-Mixed)	     
                        &(0.23, 0.26) &2.3	&1.1	&~0.41\textbf{*}	    &0.035\\
$S$ Discrete-Pure ~~(SD-Pure)     
                	    &(0.22, 0.40) &4.1	&3.3	&~0.14\textbf{*}	    &0.028\\
\hline
\end{tabular}
 \label{tab:newCRI}
 \\
\small
 {\textbf{Table 1 Mean Transits and CRI (Update the Table 3 in
 \cite{Friedman2012} with Refined Measurement)}.}\\
  *Denotes CRI Index significantly ($p$-value $< 5$\%) different from
  0 according to 2-tailed Wilcoxon test.
\end{table}

\begin{table}[h]
\center
\small
\begin{tabular}{|lcrrrrrrrc|} 
\hline
\emph{Game Condition}	&($\alpha, \beta$)&~~B$_1$$^{\dag}$	&B$_2$	&B$_3$	&B$_4$	&B$_5$	 &B$_6$& ~~~~~$\bar{C}~^{\ddag}$~~ &$p$-value 	\\
\hline
$S$ Continuous-Instant	&($\frac{1}{4},\frac{1}{4}$)  &128	&53	&120	&74	&58	&157&98.3\textbf{*}&0.028	\\
$S$ Continuous-Slow	&($\frac{1}{4},\frac{1}{4}$)  &49	&47	&45	&55	&24	&38	&43.0\textbf{*}&0.028\\
$S$ Discrete-Mixed (SD-Mixed)	&($0.23, 0.26$)  &3	&0	&3	&2	&9	&18	&5.8\textbf{*}&0.035\\
$S$ Discrete-Pure ~~(SD-Pure)	&($0.22, 0.40$)  &6	&4	&2	&4	&8	&1	&4.2\textbf{*}&0.027\\
$U_a$ Continuous-Instant&($\frac{1}{4},\frac{1}{4}$)  &158	&114	&218	&195	&175	&70&155.0\textbf{*}&0.028	\\
$U_a$ Continuous-Slow	&($\frac{1}{4},\frac{1}{4}$)  &45	&53	&48	&39	&28	&34	&41.2\textbf{*}&0.028\\
$U_a$ Discrete-Mixed	&($\frac{1}{4},\frac{1}{4}$)  &9	&16	&10	&6	&6	&17	&10.7\textbf{*}&0.027\\
$U_a$ Discrete-Pure	&($\frac{1}{4},\frac{1}{4}$)  &5	&11.5	&1.5	&13	&5.5	&11	&7.9\textbf{*}&0.028\\
$U_a$ Discrete-Pure	&($\frac{1}{4},\frac{1}{4}$)  &-52	&-29	&-43	&-49	&-40	&-33	&-41.0\textbf{*}&0.028\\
\hline
\end{tabular} \\
\textbf{Table 2: The accumulated counting number of cycles ($C$) in each block.} $^\dag$ B indicates block, number $1\sim6$ indicates the block number in the related treatment. $^\ddag$ $\bar{C}$ indicates the mean accumulated counting number of cycles of an experimental block. \textbf{*}Denotes $C$ index significantly ($p$-value $< 5$\%) different from 0 according to 2-tailed Wilcoxon test.
\label{tab:netcycle}
\end{table}

\newpage
\section*{Appendix}
CRI was defined  as \cite{Friedman2012}
    $CRI$=$\frac{CCT-CT}{CCT+CT}$,
and  $CCT$ and $CT$ can be interpreted as follows.
Supposing the Poincare section (``tripwire") is the
segment between $P_c$:$\equiv$$(\alpha, \beta)$ as shown in Fig. A1
and $P_e$=$(\alpha, 0)$,
and a transit is a directional segment from
 state ($x_1, y_1$) observed at $t$ to state ($x_2, y_2$). These two segments could
cross at $X$ as
\begin{equation}\label{eq:Crosses}
     X:\equiv(X_x, X_y) =\left(\alpha, y_1 + \frac{(y_2 - y_1)(\alpha - x_1)}{x_2 - x_1}\right).
\end{equation}
Accordingly, $CCT=\sum_{C_t>0} C_t$  and $CT=\sum_{C_t<0} |C_t|$
in which  the $C_t$ value of the transit (at time $t$) should be \footnote{For a careful description for this measurement,
see the Eq.(2) for the accumulated counting number $C$ for cycles in \cite{2013arXiv1301.3238X}. We thank
the authors' \cite{Friedman2012} confirmation on this point.}\\
\begin{tabular}{lll}
\hline
~~~~  Condition 1 ~~~~~~~~~~ & $C_t=0$ &if  $  X  \not\in  (P_c, P_e] \cup x_2 = x_1 $\\
~~~~  Condition 2  &$C_t=1$&if $  X  \in  (P_c, P_e] \cap x_2 > \alpha > x_1 $\\
   &$C_t=-1$&if $   X  \in (P_c, P_e] \cap x_2 < \alpha < x_1 $\\
~~~~  Condition 3 &$C_t=\frac{1}{2}$&if $  X \in  (P_c, P_e] \cap x_2 > x_1 \cap (x_1 = \alpha \cup x_2 = \alpha) $\\
   &$C_t=-\frac{1}{2}$&if $  X \in  (P_c, P_e] \cap x_2 < x_1 \cap  (x_1 = \alpha \cup x_2 = \alpha)$ ~~~  \\
\hline
\end{tabular} \\
  At the same time,  the accumulated counting number of cycles is $C$:=$\sum C_t$ (exactly the same as Eq.(2) in \cite{2013arXiv1301.3238X}).   According to \cite{2013arXiv1301.3238X}, when $C$ serves as an index, the criterion for cycles existence is that: if $C$ is significantly different from 0, cycles exist; alternatively, no cycle. 

 There are two mathematical mistakes in the
 measurement algorithm (see Result\_3.do in their paper's supplement).
Mistake-1: For Condition 2, a necessary condition for $C_t=\pm1$ were set
as $\frac{y_1 + y_2}{2} < \beta$ without Eq.(\ref{eq:Crosses}).
 Mistake-2:  For Condition 3, $C_t$ were set as 0.

\begin{table}[h]
\center
\small
\begin{tabular}{|lccccc|c}
\hline
\emph{~~~~~~~~Game } &&\emph{Algorithm}	    &\emph{Index}	&\emph{Mean }	& \\
\emph{~~~~~Condition}	&($\alpha, \beta$)&\emph{refined}	&\emph{chosed}	&\emph{Index}	
&~~\emph{p}-value\\
                	&&\emph{(before/after)}          &\emph{CRI/C}	   &\emph{value}  	& \\
\hline
$S$ Discrete-Pure (SD-Pure)     &(0.22, 0.40) &before	&CRI	&0.07~~ 	    &0.173\\
                	    &(0.22, 0.40) &after	&CRI	&0.14\textbf{*}	    & 0.028\\
&(0.22, 0.40) &before	&C	&  2.17~~	    & 0.113\\
                	              &(0.22, 0.40) &after	&C	&4.2\textbf{*}	 &0.027\\ 
\hline
\end{tabular}
 \label{tab:newCRI}
 \\
\small
 {\textbf{Table A1: Explanation of the Necessary of the Correction for the Measurement Algorithm}.}\\ 
  *Denotes CRI Index significantly ($p$-value $< 5$\%) different from
  0 according to 2-tailed Wilcoxon test.
\end{table}

 \begin{figure}
  \begin{center}
    \includegraphics[width=.45\linewidth]{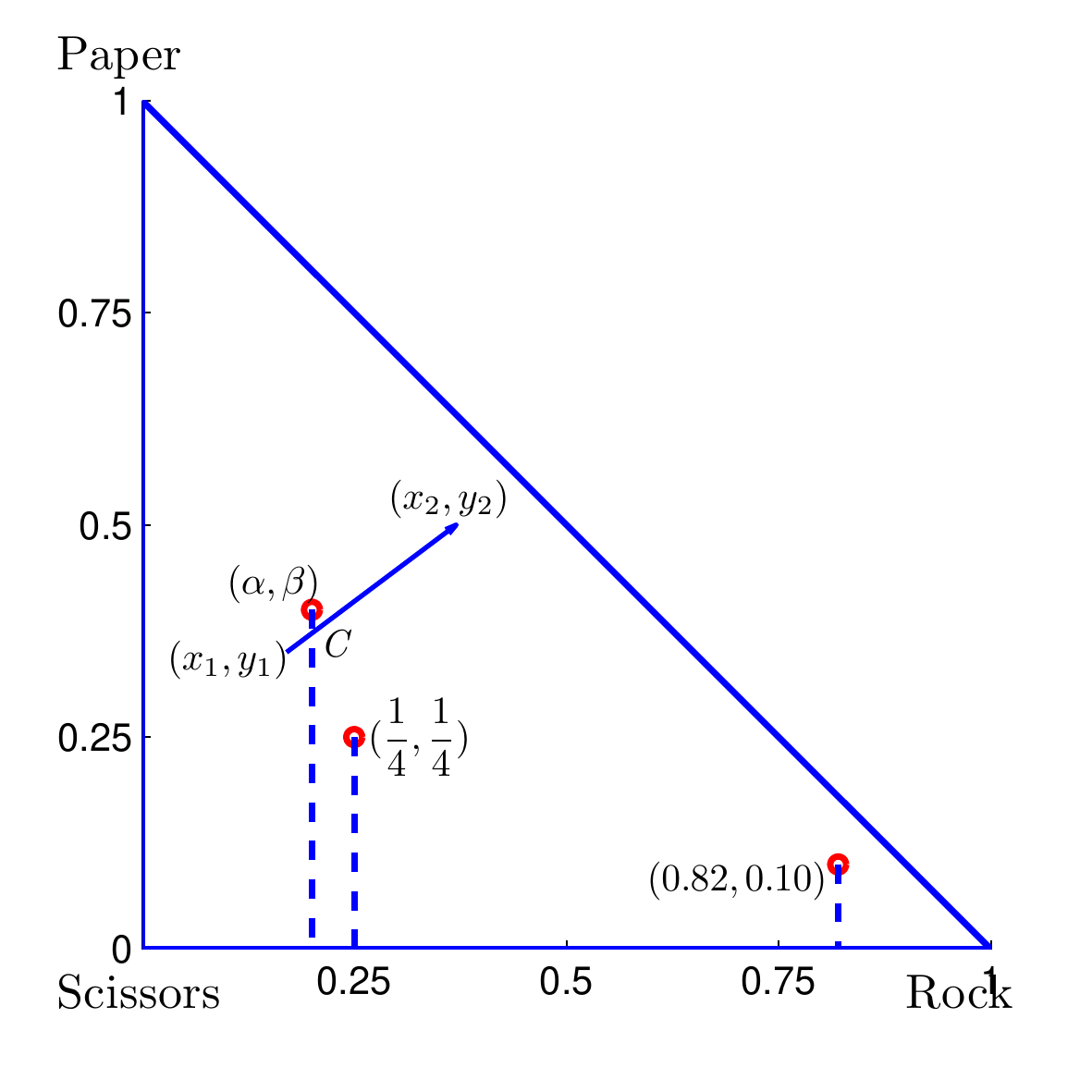}
  \end{center}
      \small
    \textbf{Figure A1: Illustration of the tripwire for cycles counting (Poincare section)\cite{Friedman2012}.}
      A tripwire for cycles counting is a line segment constructed between the reference point $(\alpha, \beta)$
      and the simplex edge, as the vertical dashed line extending below the point $(\alpha, \beta)$.
      The right tripwire is set as the reference point $(\alpha, \beta)$=(0.82, 0.10).
      The authors \cite{Friedman2012} set $(\alpha, \beta)=(\frac{1}{4}, \frac{1}{4})$
      (the Nash equilibrium), used the tripwire (middle) and observed weak and no cycles for stable discrete time treatments (SD).
      The arrow indicates a transit from $(x_1,y_1)$ to $(x_2,y_2)$ crossing the left tripwire at $C$.
\end{figure}

These mistakes need to be corrected. Table A1 exhibits both results, without algorithm refined and with algorithm refined.  Obviously, without refined algorithm we cannot find cycles even if we set $(\alpha, \beta)$ at $(0.22, 0.40)$. On contrary, with the refined algorithm, the existence of cycles can be verify by both $CRI$ and $C$. This comparison implies that the algorithm correction is necessary.

 Programs for the replication of the results in this paper are provided as supplements.

 \section*{Acknowledgments} We thank John Ledyard, Charles Plott, Hai-Jun Zhou, Daniel Friedman and Qiqi Cheng for helpful discussion. This work was supported by Grants from 985 at Zhejiang University, SKLTP of ITP CAS (No.Y3KF261CJ1) and Philosophy \& Social Sciences Planning Project of Zhejiang Province (13NDJC095YB).

 \bibliographystyle{jpe}

%
%



\end{document}